\documentclass[10pt]{article} % For LaTeX2e
\usepackage[preprint]{tmlr}
\usepackage{amsmath,amsfonts,bm}

\def\eqref#1{equation~\ref{#1}}
\def\1{\bm{1}}

\DeclareMathAlphabet{\mathsfit}{\encodingdefault}{\sfdefault}{m}{sl}
\SetMathAlphabet{\mathsfit}{bold}{\encodingdefault}{\sfdefault}{bx}{n}

\usepackage{hyperref}
\usepackage{url}
\usepackage{listings}
\usepackage{graphicx}
\usepackage{float}

\definecolor{verylightgray}{gray}{0.95}

\title{BPD-Neo: An MRI Dataset for Lung-Trachea Segmentation with
Clinical Data for Neonatal Bronchopulmonary Dysplasia}

\author{\name Rachit Saluja \email rs2492@cornell.edu \\
\addr Cornell University \& Cornell Tech\\
Weill Cornell Medicine
\AND
\name Arzu Kovanlikaya  \\
\addr Weill Cornell Medicine
\AND
\name Candace Chien \\
\addr Weill Cornell Medicine 
\AND
\name Lauren Kathryn Blatt \\
\addr Weill Cornell Medicine 
\AND
\name Jeffrey M. Perlman \\
\addr Weill Cornell Medicine 
\AND
\name Stefan Worgall \\
\addr Weill Cornell Medicine 
\AND
\name Mert R. Sabuncu\textsuperscript{*}  \\
\addr Cornell University \& Cornell Tech \\
Weill Cornell Medicine 
\AND
\name Jonathan P. Dyke\textsuperscript{*} \textsuperscript{\dag} \email jpd2001@med.cornell.edu \\
\addr Weill Cornell Medicine 
}

\def\month{MM}  % Insert correct month for camera-ready version
\def\year{YYYY} % Insert correct year for camera-ready version
\def\openreview{\url{https://openreview.net/forum?id=XXXX}} % Insert correct link to OpenReview for camera-ready version

\begin{document}

\maketitle

\begin{abstract}
Bronchopulmonary dysplasia (BPD) is a common complication among preterm neonates, with portable X-ray imaging serving as the standard diagnostic modality in neonatal intensive care units (NICUs). However, lung magnetic resonance imaging (MRI) offers a non-invasive alternative that avoids sedation and radiation while providing detailed insights into the underlying mechanisms of BPD. Leveraging high-resolution 3D MRI data, advanced image processing and semantic segmentation algorithms can be developed to assist clinicians in identifying the etiology of BPD. In this dataset, we present MRI scans paired with corresponding semantic segmentations of the lungs and trachea for 40 neonates, the majority of whom are diagnosed with BPD. The imaging data consist of free-breathing 3D stack-of-stars radial gradient echo acquisitions, known as the StarVIBE series. Additionally, we provide comprehensive clinical data and baseline segmentation models, validated against clinical assessments, to support further research and development in neonatal lung imaging.

\vspace{5em}
\noindent\rule{0.5\textwidth}{0.4pt}
\\
\textsuperscript{*} Contributed equally as senior co-authors.
\\
\textsuperscript{\dag} Corresponding Author.

\end{abstract}

\newpage

\section{Background \& Summary}

Automated segmentation of the neonatal respiratory system is particularly relevant for preterm newborns at risk of developing bronchopulmonary dysplasia (BPD). The etiology of BPD manifests through multiple mechanisms, including tracheobronchomalacia, parenchymal lung disease, pulmonary hypertension, or a combination of these factors \citep{wu2020characterization}. While portable X-ray remains the standard imaging modality in neonatal intensive care units (NICUs), lung MRI offers a non-invasive alternative that provides detailed insights into these pathological mechanisms without the need for sedation or radiation \citep{dyke2023assessment, stewart2025pulmonary}. Automated segmentation of the neonatal trachea and lung volume could enhance clinicians' ability to identify the underlying causes of BPD, facilitating improved diagnosis and management.

\begin{figure}[H]
\begin{center}
\includegraphics[width=\linewidth]{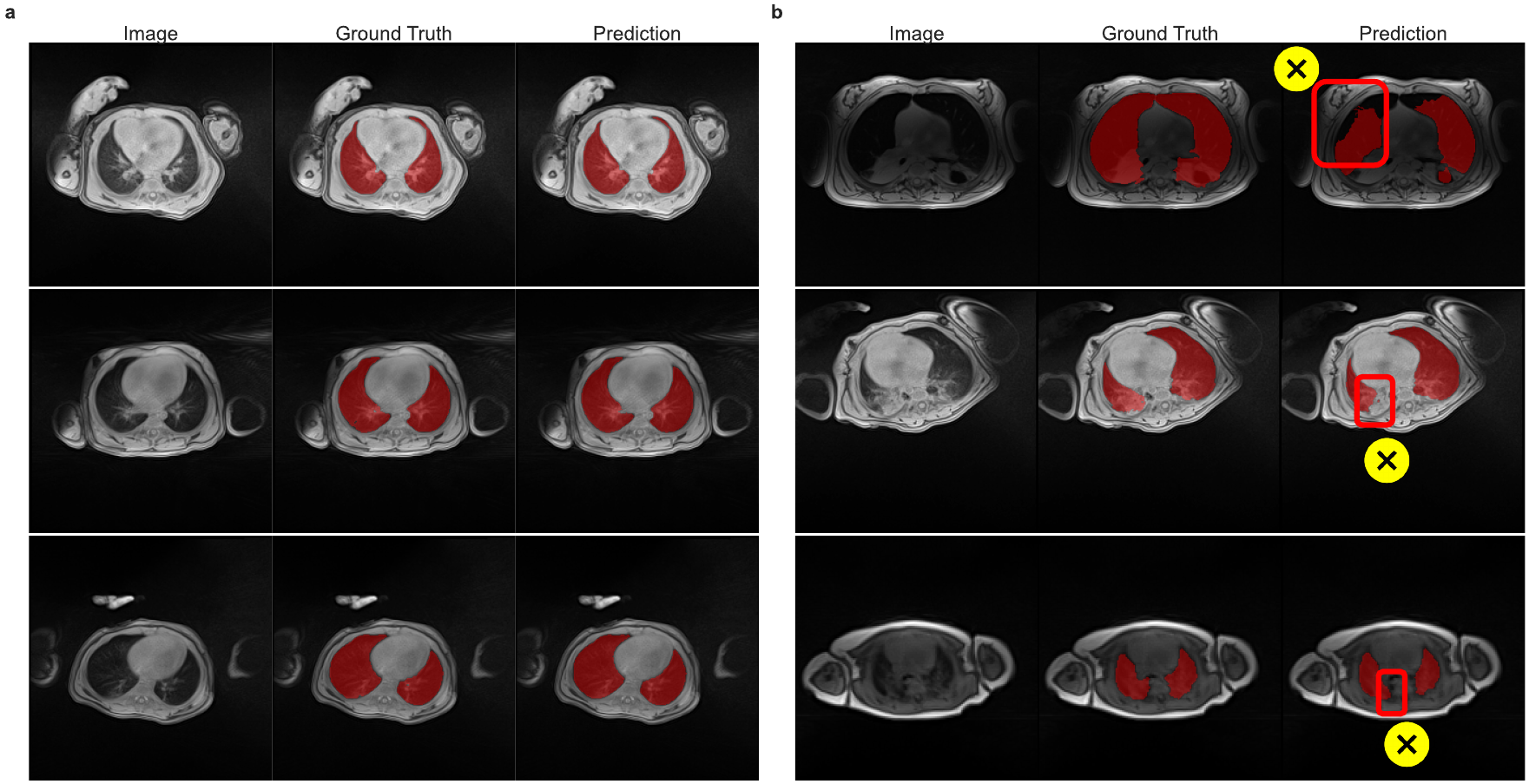}
\end{center}
\caption{Examples of T1-weighted StarVIBE images (axial view) and their corresponding ground truth and predicted segmentations of lungs (a) illustrates outputs for high-performing cases, (b) highlights cases with lower performance, including instances of both over-segmentation and under-segmentation}
\label{Figure1a}
\end{figure}

The compliant airway in preterm infants is particularly susceptible to injury due to prolonged endotracheal intubation and exposure to positive pressure ventilation. Airway malacia refers to an abnormally compliant airway, leading to excessive collapse (>50\% reduction in airway area). Notably, large airway disease is observed in approximately one-third of infants diagnosed with BPD, yet it often remains undiagnosed \citep{hysinger2017tracheobronchomalacia}. The current gold-standard diagnostic method, bronchoscopy, requires sedation and carries inherent procedural risks. Tracheobronchomalacia is further associated with increased morbidity, including prolonged hospital stays, a higher incidence of pneumonia, and an increased likelihood of requiring tracheostomy placement. The use of MRI to quantify tracheal airway area presents a promising non-invasive alternative, enabling the automatic segmentation of the airway through deep learning models for more accessible and risk-free diagnosis. Historically, parenchymal lung disease has been the defining feature of BPD. Parenchymal lung disease as a cause in BPD is made evident in both structural and functional lung MRI \citep{dyke2023assessment, higano2017quantification}. Preterm birth results in the arrest of lung development resulting in ineffective gas exchange and need for respiratory support and oxygen. Parenchymal injury is complex, resulting from multiple antenatal and postnatal exposures which further disrupt alveolarization and lead to abnormal repair. Since the introduction of antenatal steroids and surfactant, BPD is mostly characterized by a large simplified alveolar structure. More severe BPD patients have heterogeneous parenchymal disease characterized by atelectasis, hyperinflation, edema, and fibrosis. MRI may be used to quantitate the degree of parenchymal disease in the lung.

The application of semantic segmentation models for quantifying lung and tracheal volumes in MRI images remains uncommon. Recently, a study developed a model to quantify lung volumes in BPD patients; however, while that study had a larger sample size, the dataset was not publicly available, and no clinical data were included \citep{mairhormann2023automated}. Additionally, their analysis did not utilize StarVIBE MRI series, which we believe may offer greater utility for this application. \citep{mairhormann2023automated} also utilizes the BPD grading system proposed by \citep{jobe2001bronchopulmonary}. In contrast, our study adopts the more contemporary 2019 Jensen criteria (\citep{jensen2019diagnosis}), which are also used in current clinical practice and are expected to provide a more accurate and clinically relevant categorization of BPD severity.

Our dataset provides clinicians and researchers with the resources to develop semantic segmentation models capable of automatically segmenting lung and tracheal volumes. The binary image masks produced by the semantic segmentation models may be multiplied by a structural UTE or StarVIBE MRI sequence to produce a parenchymal signal intensity histogram \citep{higano2017quantification, vanhaverbeke2020lung}. The degree of hyperinflation (ratio of total-lung-volume [TLV] to body-surface-area [BSA]) may also be measured using the Mosteller formula and the derived lung volume. Automatically segmenting the neonatal respiratory system using deep learning-based semantic segmentation methods offers a rapid and objective approach for clinicians to assess tracheal and lung health in relation to the various etiologies of BPD. Our dataset is the first open-source resource to provide paired imaging and segmentation data, facilitating the development of advanced computational models for neonatal respiratory assessment.

\begin{figure}[H]
\begin{center}
\includegraphics[width=\linewidth]{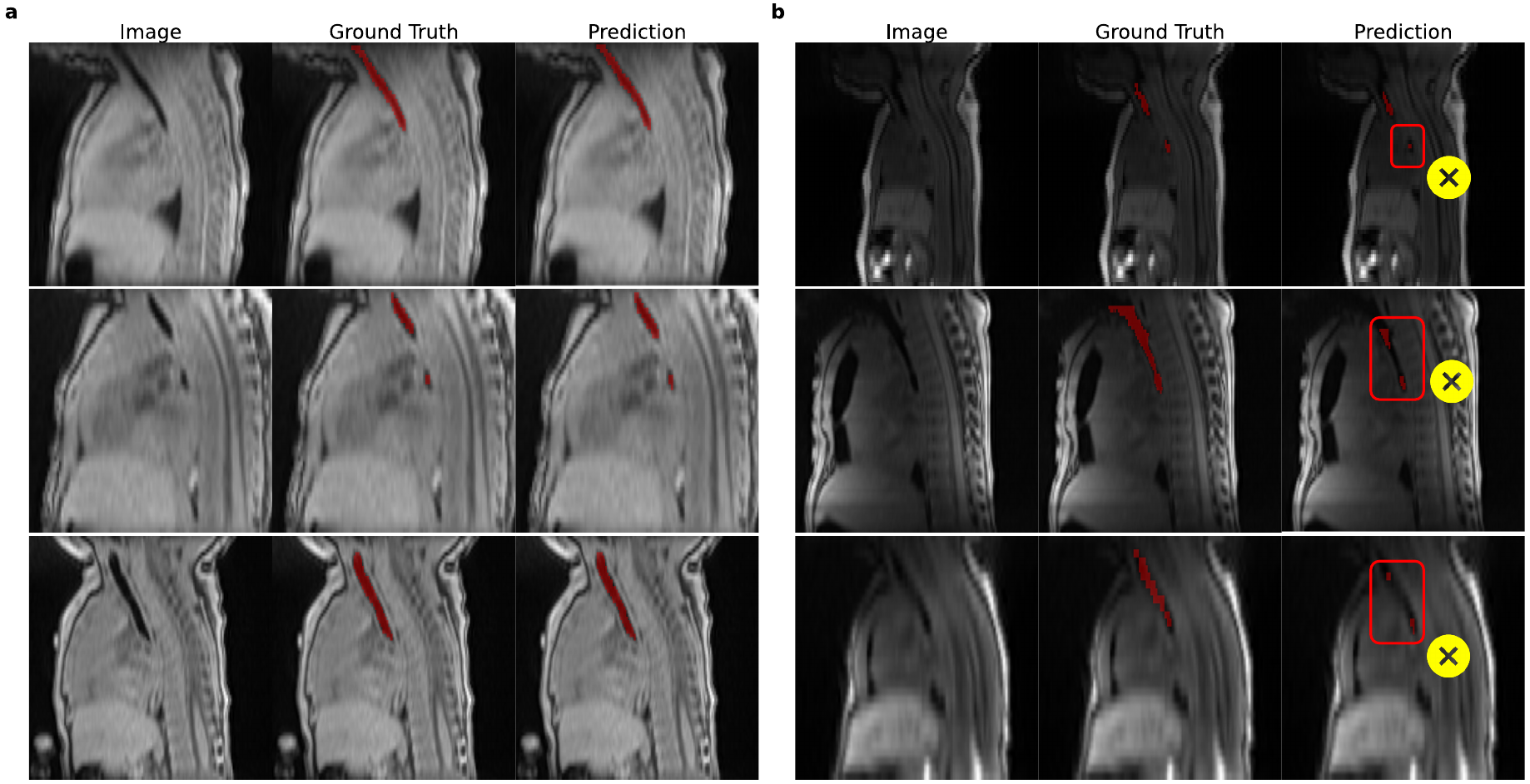}
\end{center}
\caption{Examples of T1-weighted StarVIBE images (sagittal view) and their corresponding ground truth and predicted segmentations of trachea (a) illustrates outputs for high-performing cases, (b) highlights cases with lower performance, including instances of both over-segmentation and under-segmentation}
\label{Figure1b}
\end{figure}

\section{Methods}

\subsection{Data Collection}

Neonatal lung volume was assessed from 40 neonates (18 M/22 F) enrolled in a prospective study assessing physiologic phenotyping of chronic lung disease of prematurity using MRI (NHLBI; R01-HL167003). All parents signed informed consent as part of an approved protocol of our Institutional Review Board. Inclusion criteria were infants less than 6 month of age in the NICU already receiving a clinically indicated brain MRI as part of standard of care. An additional 15 minutes was added at the completion of the brain MRI to perform the research lung imaging sequences without the administration of any contrast agents.

\subsection{MRI Data Acquisition}

MRI data was acquired on a 1.5 Tesla Siemens Amira Scanner (Siemens Healthineers; Erlangen, Germany) located in the Neonatal Intensive Care Unit of the New York-Presbyterian Alexandra Cohen Hospital for Women and Newborns. While the availability of MRI within the NICU is a unique feature of our institution which enhances accessibility and comfort for both the infant and family, similar scans can feasibly be conducted outside the NICU, potentially increasing their clinical applicability and broader availability. Infants were fed, swaddled and transported within the NICU to the MRI scanner. Multiple layers of hearing protection were employed to minimize the acoustic noise reaching the infant. Earplugs were used which in general can reduce the noise in the MRI by between 20 and 30 decibels (dB) and are always the first line of defense. In addition to standard foam earplugs, we used MRI safe disposable MiniMuff neonatal noise guards (Natus Medical, San Carlos, CA, USA) which have a gentle hydrogel adhesive to provide a secure fit and to reduce the noise by an additional 7dB. Noise canceling infant MRI headphones were lastly used (Ima-X; Luxembourg) which reduced the noise by an average of 22 dB with up to 30 dB @ 1kHz. During the MRI study, the infant’s vitals were monitored continuously by a neonatal nurse using a Philips MR400 patient monitoring system with infant accessories. The infant was also audibly monitored for any signs of distress and the scan immediately stopped should the infant experience any discomfort, and the scan not restarted until they were calmed. 

A pair of 8-channel NORAS VARIETY flex coils (20 cm x 22 cm) (NORAS MRI products, Höchberg, Germany), were used to provide one coil anterior and one coil posterior on the infant. A free-breathing 3D stack-of-stars radial gradient echo technique known as StarVIBE was acquired axially for segmentation of both lung and trachea in the neonates \citep{block2013improving, azevedo2011free}. StarVIBE is optimally used in pediatric patients and is robust in resisting motion artifacts. Specific acquisition parameters included a 20 cm field of view (FOV) and a 224 x 224 matrix size yielding a 0.9 mm x 0.9 mm x 2 mm (1.6 ml) voxel resolution. A repetition time (TR) of 4.2 ms, an echo time (TE) of 2.0 ms, a flip angle of 4º and a receive bandwidth of 603 Hz/pixel were used. 

\begin{table}[t]
\begin{center}
\begin{tabular}{ll}
\multicolumn{1}{c}{\bf Clinical Data}  &\multicolumn{1}{c}{\bf Description}
\\ \hline \\
Weight (grams)   &Weight of premature infant at MRI scan in grams \\
Length (cm)      &Length of premature infant at birth in cms \\
Sex              &Sex of premature infant \\
BW (grams)       &Weight of premature infant at the time of the MRI in grams \\
GA (weeks)       &Gestational age in weeks \\
PMA at Study (weeks) &Postmenstrual age in weeks \\
Jensen 2019 BPD Definition & Premature infant's Jensen 2019 BPD Classification
\end{tabular}
\caption{Full list of clinical data and their description}
\label{clinical Data}
\end{center}
\end{table}

\subsection{Expert Image Annotation}

Segmentation of the lungs and trachea was conducted using 3D Slicer \url{http://slicer.org} and its segmentation editor \citep{fedorov20123d, pinter2019polymorph}. Fiducial markers (seed points) were manually placed at the center of each lung and tracheal slice, followed by application of a region-growing algorithm to delineate the structures. A smoothing kernel of 3 mm × 3 mm × 1 mm was applied to the lung segmentations to refine boundaries, whereas no smoothing was applied to the tracheal segmentations due to the limited voxel count in those regions of interest. Final segmentations were reviewed, manually corrected for any errors, and validated by a domain expert before being exported as NIfTI files. Additionally, we provide a small subset of expert segmentations from a second reviewer to facilitate inter-observer variability analysis, as detailed in Section \ref{IO-section}. This reviewer followed the exact same segmentation methodology to ensure consistency in the annotation process.

\subsection{Clinical Data}

In addition to the imaging data, key clinical variables were collected, including birth weight, weight at the time of MRI, gestational age, and postmenstrual age, along with BPD classification based on the Jensen 2019 criteria. The availability of these clinical data alongside imaging data facilitates future research into the identification of biomarkers derived from segmentation-based volumetrics, enabling their integration with clinical variables to enhance understanding of disease progression and outcomes. A comprehensive list of the clinical variables collected and included in the dataset is presented in Table \ref{clinical Data}.

\section{Data Records}

All data records, including the DICOM series, NIfTI files, and clinical data, are available at \url{https://zenodo.org/records/15768091}. The dataset includes a XLSX file containing the clinical data, matched to each study by study identifier. The DICOM data comprise all imaging series acquired during the MRI sessions, have been fully anonymized, and are suitable for future research applications. Additionally, the dataset contains NIfTI files for the lung and trachea segmentations corresponding to each study along with some of the multi-rater segmentations. Below is the directory structure of the data record:

\begin{lstlisting}
.
|-- clinical_data.xlsx                            ## Clinical Data
|-- DICOM-data/                                   ## Dicom Data
|   |-- BPD-Neo-01/                               ## Study
|   |   |-- SER0001/                              ## Dicom Series 
|   |   |   |-- IMG00001.dcm
|   |   |   |-- IMG00002.dcm
|   |   |   |-- ...
|   |   |   `-- IMG00018.dcm
|   |   |-- SER0002/
|   |   |-- ...
|   |   `-- SER0007/
|   |-- BPD-Neo-02/
|   |-- ...
|   `-- BPD-Neo-40/
`-- Nifti-data/
    |-- BPD-Neo-01/                               ## Nifti Data
    |   |-- image.nii.gz                          ## Image
    |   |-- lung_seg.nii.gz                       ## Lung Segmentation
    |   `-- trachea_seg.nii.gz                    ## Trachea Segmentation
    |-- ...
    `-- BPD-Neo-40/
\end{lstlisting}

\newpage

\begin{figure}[H]
\begin{center}
\includegraphics[width=\linewidth]{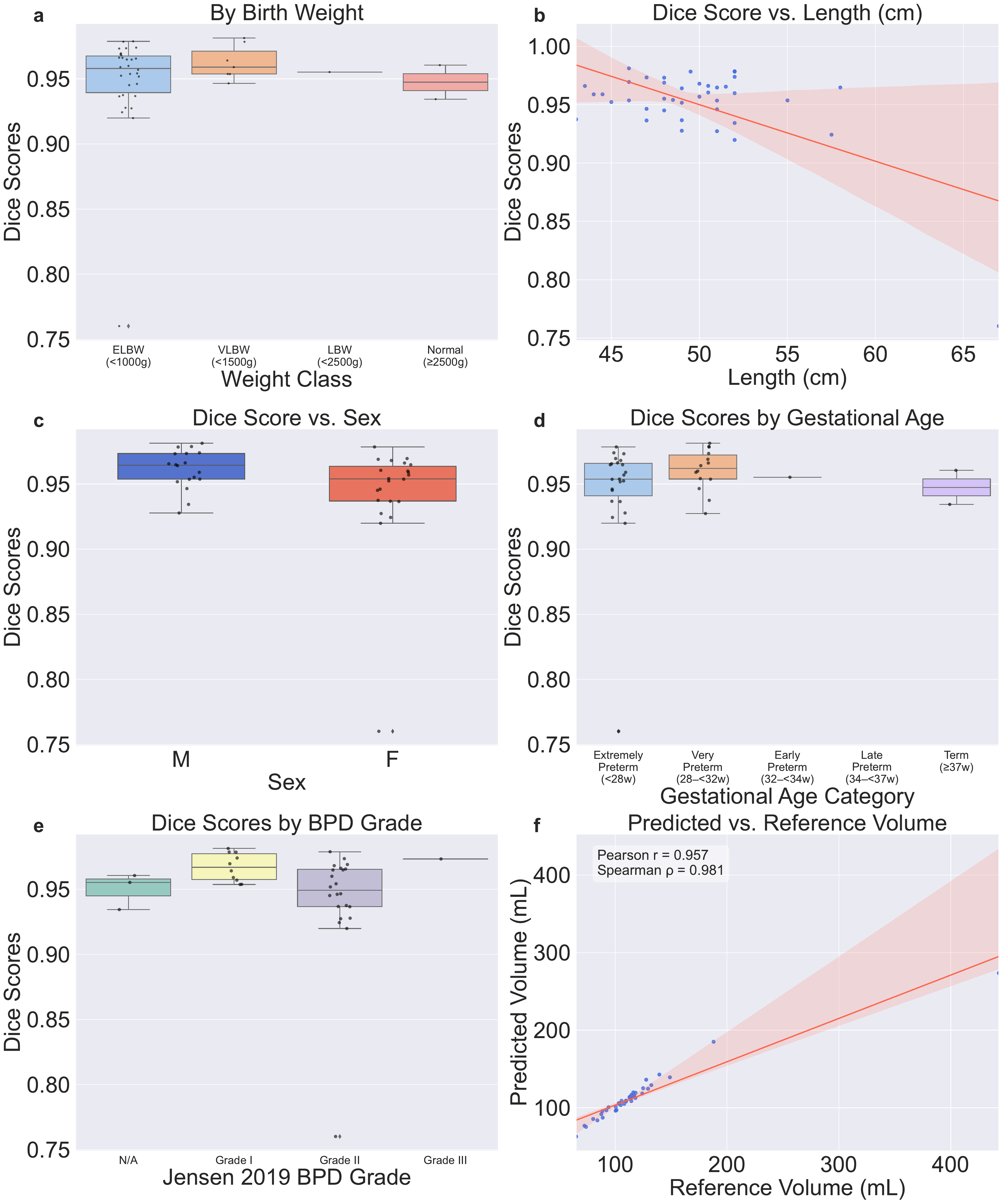}
\end{center}
\caption{Evaluation of lung segmentation model against clinical variables.}
\label{Figure2}
\end{figure}

\newpage

\begin{figure}[H]
\begin{center}
\includegraphics[width=\linewidth]{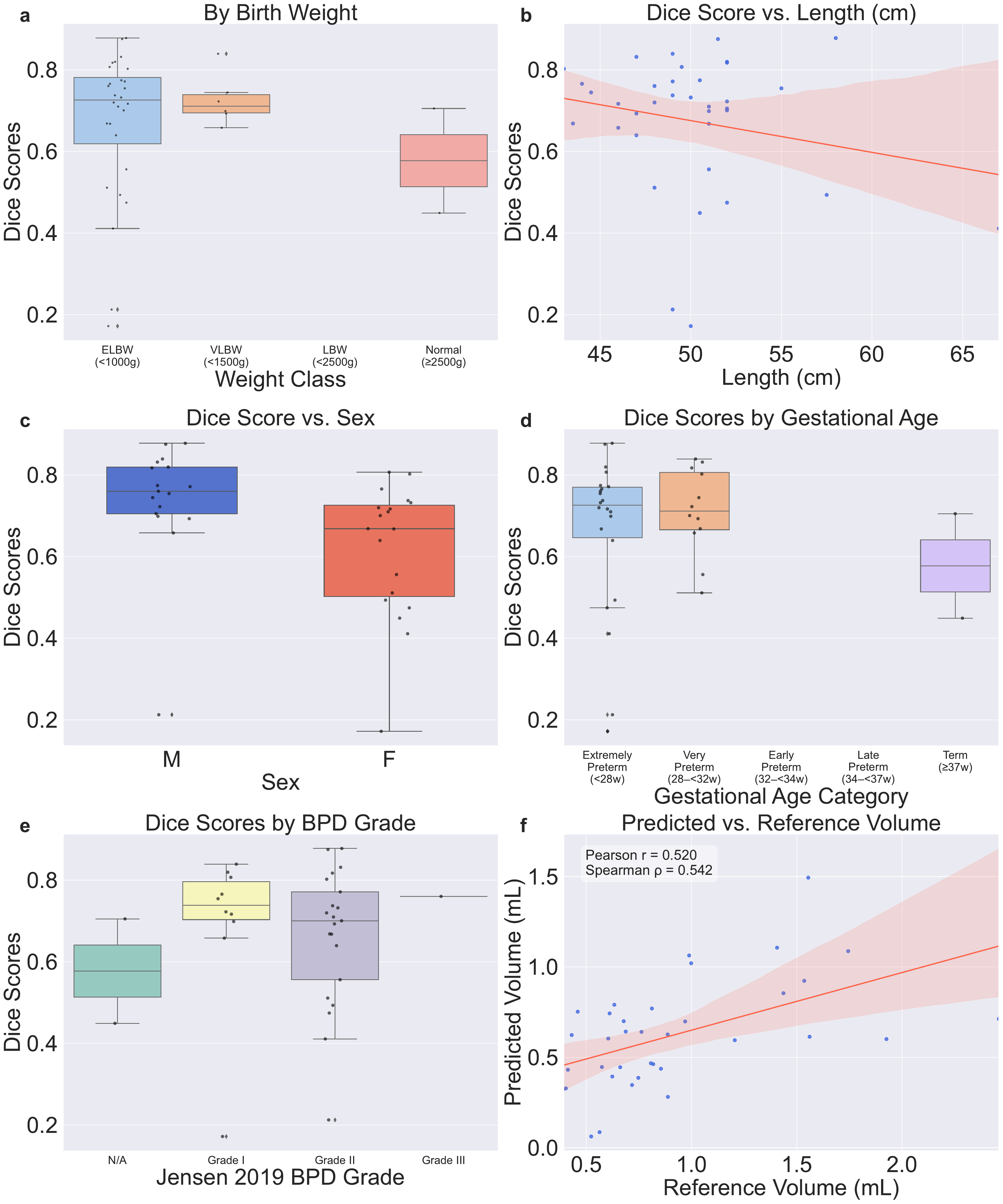}
\end{center}
\caption{Evaluation of trachea segmentation model against clinical variables.}
\label{Figure3}
\end{figure}

\newpage

\section{Technical Validation}

\subsection{Segmentation Model}

To support further scientific research, we benchmark our dataset by providing pretrained segmentation models. Specifically, we train two U-Net models \citep{ronneberger2015u} for lung and trachea segmentation using the nnUNetV2 framework \citep{isensee2021nnu} on T1-weighted StarVIBE MRI data. The models are based on the 3D full-resolution U-Net architecture, optimized to accurately delineate the lungs and trachea through a combination of soft Dice loss and cross-entropy loss with deep supervision.

The lung segmentation model was trained on a dataset comprising 40 samples, partitioned into five folds using an 80/20 split for training and validation. The trachea segmentation model was trained on 36 samples using the same 5-fold cross-validation strategy to ensure robustness and generalizability. Additionally, we conducted an ablation study by replacing the U-Net with residual connections in the encoder \citep{isensee2024nnu} to evaluate the impact of architectural modifications on segmentation performance.

\begin{table}[h]
\centering
\begin{tabular}{|l|l|c|c|c|c|c|c|c|c|c|}
\hline
\textbf{Model} & \textbf{Dataset} & \textbf{N} & \textbf{Fold-0} & \textbf{Fold-1} & \textbf{Fold-2} & \textbf{Fold-3} & \textbf{Fold-4} & \textbf{Mean} & \textbf{Std} \\
\hline
nnUNet & BPD-Neo-Lung     & 40 & 0.961 & 0.955 & 0.952 & 0.932 & 0.956 & 0.9512 & 0.0112 \\
nnUNet(ResEnc) & BPD-Neo-Lung & 40 & 0.957 & 0.954 & 0.952 & 0.928 & 0.950 & 0.9482 & 0.0116 \\
nnUNet & BPD-Neo-Trach  & 36 & 0.705 & 0.635 & 0.665 & 0.765 & 0.599 & 0.6738 & 0.0642 \\
nnUNet(ResEnc) & BPD-Neo-Trach & 36 & 0.672 & 0.662 & 0.598 & 0.740 & 0.626 & 0.6596 & 0.0537 \\
\hline
\end{tabular}
\caption{Cross-validation dice score performance of nnUNet and nnUNet (ResEnc) models}
\label{cv-results}
\end{table}

The cross-validation performance results are presented in Table~\ref{cv-results}. The lung segmentation model demonstrates strong delineation capabilities, achieving a high mean cross-validation Dice score of 95.1\%, indicating reliable performance which is reflected in Figure~\ref{Figure1a}. In contrast, the trachea segmentation model yielded a lower peak Dice score of 67.3\%, suggesting that further refinement is needed to improve performance in this task, as shown in Figure~\ref{Figure1b}. One potential factor contributing to the reduced accuracy is the placement of NORAS Flex coils over the lungs, which may have led to a decline in signal intensity, in turn reducing the performance when segmenting the trachea, particularly as the distance from the coils increased.

\subsection{Segmentation Performance Versus Clinical Data}

To assess the clinical relevance of our segmentation models, we evaluated the relationship between Dice scores, segmentation-derived volumes, and corresponding clinical variables. For this analysis, we utilized the nnUNet models rather than the nnUNet (ResEnc) variants, due to their superior performance. As shown in Figure~\ref{Figure2} panels (a), (c), (d), and (e), the models achieved consistently high Dice scores across different birth weight classes, sex, gestational age categories, and Jensen 2019 BPD grades, indicating robust and generalizable performance across clinically relevant subgroups. 

As shown in Figure~\ref{Figure2} panel (f), we also observe a strong correlation between the predicted and reference lung volumes, with a Pearson correlation coefficient of 0.957 and a Spearman correlation coefficient of 0.981, indicating high agreement between model outputs and expert annotations. Additionally, we note a slight decrease in Dice score, dropping to approximately 0.92, as infant length increases, suggesting that anatomical variability associated with body size may modestly affect segmentation performance.

\begin{table}[h]
\centering
\begin{tabular}{|l|l|c|c|}
\hline
\textbf{Target} & \textbf{N} & \textbf{R1 vs R2} (Dice) & \textbf{$\text{R}_{staple}$ vs Best Model} (Dice)\\
\hline
Lung & 11 & 0.955 $\pm$ 0.014 & 0.940 $\pm$ 0.050 \\
Trachea & 07 & 0.734 $\pm$ 0.099 & 0.676 $\pm$ 0.120  \\
\hline
\end{tabular}
\caption{Inter-reviewer and model performance comparison for lung and trachea segmentation. The mean Dice similarity coefficient ($\pm$ standard deviation) is reported for both (i) inter-rater agreement between Reviewer 1 (R1) and Reviewer 2 (R2), and (ii) agreement between the STAPLE-generated consensus segmentation ($\text{R}_{staple}$) and the best-performing model. Agreement is notably higher for lung segmentation compared to trachea, both between human raters and between model and consensus.}
\label{io-results}
\end{table}

For tracheal segmentation, the overall Dice performance was lower, and greater variability was observed in relation to clinical variables. As illustrated in Figure~\ref{Figure3}, Dice scores varied across different birth weight classes, sex, Jensen 2019 BPD grades, and gestational age categories, indicating sensitivity to clinical and anatomical heterogeneity. Additionally, a lower correlation was observed between predicted and reference tracheal volumes, along with increased variation in Dice scores as a function of infant length, suggesting that tracheal segmentation may be more susceptible to anatomical and imaging variability. Future studies may focus on developing higher-performing models for tracheal segmentation, aiming to improve robustness and accuracy in this challenging subset of the dataset.

\begin{figure}[h]
\begin{center}
\includegraphics[width=\linewidth]{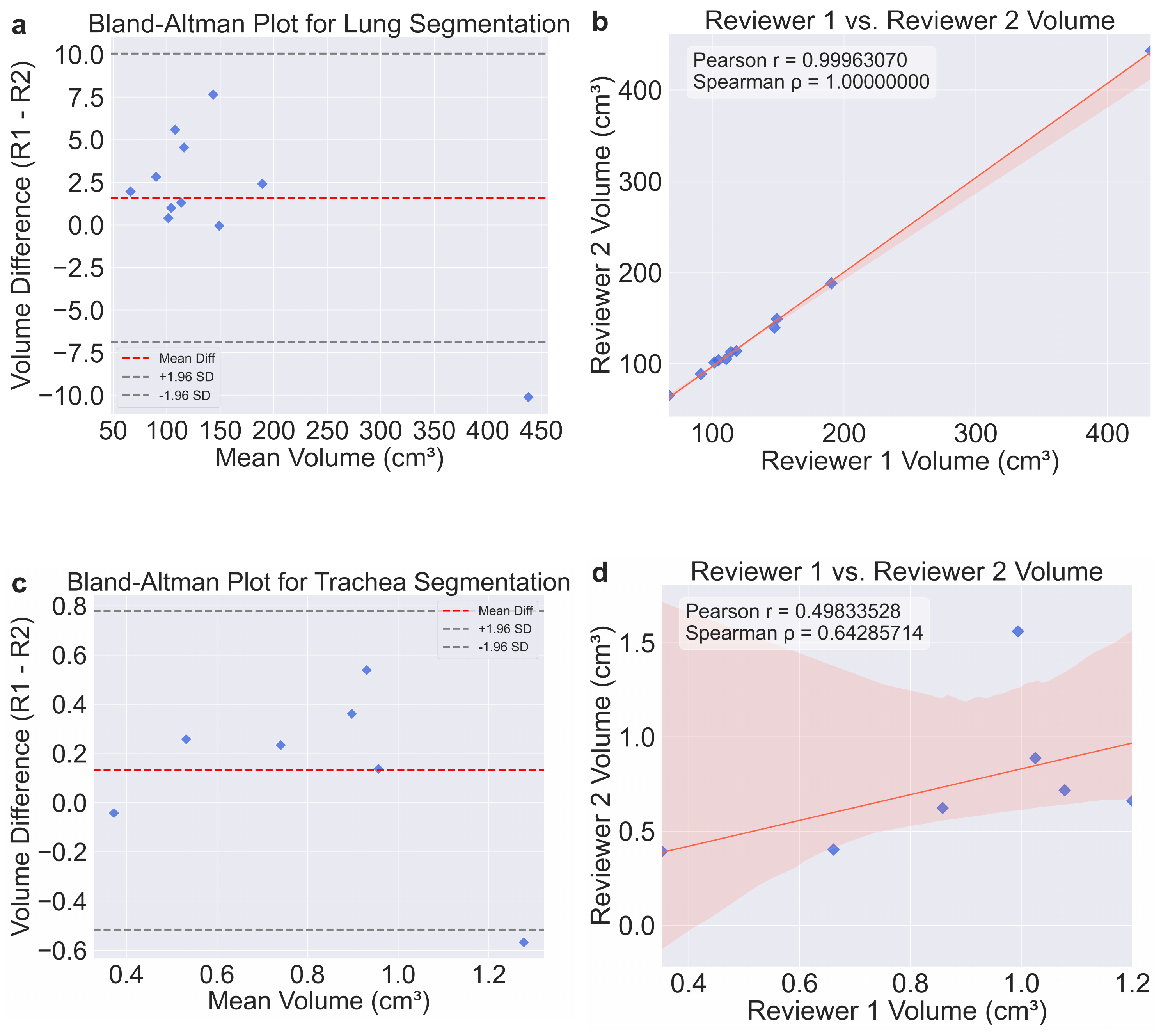}
\end{center}
\caption{Inter-rater agreement for lung and trachea segmentation volumes.
(a, c) Bland–Altman plots comparing volume differences between Reviewer 1 and Reviewer 2 for lung (a) and trachea (c) segmentations. The red dashed line indicates the mean difference, while the gray dashed lines denote the $\pm$1.96 standard deviation limits of agreement.
(b, d) Scatter plots with linear regression comparing segmentation volumes between the two reviewers for lung (b) and trachea (d). Pearson correlation coefficients and Spearman correlation coefficients are reported. Agreement is high for lung segmentation (Pearson r $\approx$ 1.0), whereas trachea segmentation shows weaker correlation and greater variability.
}
\label{Figure5}
\end{figure}

\subsection{Inter-Observer Variability}
\label{IO-section}

We also assess inter-observer variability for both segmentation tasks, as it provides an important upper bound for model performance and highlights the inherent variability among human reviewers. This benchmark helps contextualize the model’s accuracy relative to expert-level agreement and establishes a practical ceiling for achievable performance. 

To evaluate inter-observer variability, a second expert independently segmented a subset of the dataset, Lungs (N = 11) and Trachea (N = 7), following the same segmentation protocol as the first observer. This included identical preprocessing steps, fiducial placement, region-growing procedures, and post-processing corrections, ensuring consistency in the annotation methodology for comparative analysis.

We first compare the segmentation performance metrics between Reviewer 1 and Reviewer 2. The inter-observer Dice score for lung segmentation was 0.955 $\pm$ 0.014 (from Table \ref{io-results}), indicating very high agreement and serving as a pseudo-upper bound for model performance. Notably, our best model achieved a mean 5-fold cross-validation Dice score of 0.9512, demonstrating that the model performs comparably to expert-level consistency and is well-optimized for the task.

From Table~\ref{io-results}, we also observe that inter-observer Dice scores for trachea segmentation are substantially lower compared to lung segmentation, with a mean score of 0.734 $\pm$ 0.099. This indicates greater variability and disagreement between reviewers for tracheal annotations, likely due to the smaller structure size and lower signal quality. This inter-observer variability is reflected in the model’s performance, with the best-performing trachea segmentation model achieving a 5-fold cross-validation Dice score of 0.6738.

To further analyze inter-observer variability, we computed Bland-Altman plots and measured the volume correlation coefficient between the two reviewers. As shown in Figure \ref{Figure5}(a), the majority of differences in lung volumes fall within a narrow range of approximately -10 to +10 $\text{cm}^3$, indicating strong agreement for lung segmentation. In contrast, Figure \ref{Figure5}(c) reveals substantially greater variability in tracheal volume estimates relative to their mean, which is expected given the smaller absolute volumes and increased difficulty in delineating the trachea.

Figure \ref{Figure5} panels (b) and (d) illustrate the volume correlations between reviewers. As shown in panel (b), there is strong agreement for lung segmentation, with the reviewers' measurements closely aligned. However, panel (d) highlights substantially greater variability in trachea segmentation, underscoring the challenges associated with accurately delineating smaller airway structures. This analysis indicates that there remains considerable room for improvement and further research in trachea segmentation for neonatal StarVIBE MR images.

Finally, we generated a consensus segmentation for both lung and trachea using the STAPLE algorithm \citep{warfield2004simultaneous}, integrating the annotations from both reviewers. We then computed the Dice scores between this consensus and the predictions from the best-performing model. As shown in Table \ref{io-results}, the trend remains consistent, high Dice scores were achieved for lung segmentation, while lower scores were observed for trachea segmentation. This further supports the notion that tracheal segmentation remains a more challenging task and highlights the need for continued research in this area.

\subsubsection*{Usage Notes}
Users should cite this paper in their research output and acknowledge the contribution of this dataset in their study.

\subsubsection*{Code availability}
The code repository for the segmentation models can be accessed via \url{https://github.com/rachitsaluja/BPD-Neo}. 

\subsubsection*{Author Contributions}

All authors meet the following criteria:

\begin{enumerate}
    \item Substantial contributions to the conception or design of the work; or the acquisition, analysis, or interpretation of data for the work
    \item Drafting the work or reviewing it critically for important intellectual content
    \item Final approval of the version to be published
    \item Agreement to be accountable for all aspects of the work in ensuring that questions related to the accuracy or integrity of any part of the work are appropriately investigated and resolved.
\end{enumerate}

\subsubsection*{Acknowledgments}
The authors would like to acknowledge the assistance of the pediatric and NICU nursing staff at NYP who were invaluable in the success of this study. Funding for this work is provided under NHLBI R01HL167003.  

\bibliography{main}
\bibliographystyle{tmlr}

% \appendix
% \section{Appendix}

\end{document}